# HNote: Extending YNote with Hexadecimal Encoding for Fine-Tuning LLMs in Music Modeling

Hung-Ying Chu[1], Shao-Yu Wei[1], Guan-Wei Chen[1], Tzu-Wei Hung[1], Cheng-Yang Tsai[1] and Yu-Cheng Lin[1]

[1] Department of Computer Science and Engineering, Yuan Ze University, Taoyuan, Taiwan
{ s1111521,s1123332,s1111446,s1113315,s1111505}@mail.yzu.edu.tw
linyu@saturn.yzu.edu.tw

**Abstract.** Recent advances in large language models (LLMs) have created new opportunities for symbolic music generation. However, existing formats such as MIDI, ABC, and MusicXML are either overly complex or structurally inconsistent, limiting their suitability for token-based learning architectures. To address these challenges, we propose HNote, a novel hexadecimal-based notation system extended from YNote, which encodes both pitch and duration within a fixed 32-unit measure framework. This design ensures alignment, reduces ambiguity, and is directly compatible with LLM architectures. We converted 12,300 Jiangnan-style songs generated from traditional folk pieces from YNote into HNote, and fine-tuned LLaMA-3.1(8B) using parameter-efficient LoRA. Experimental results show that HNote achieves a syntactic correctness rate of 82.5%, and BLEU and ROUGE evaluations demonstrate strong symbolic and structural similarity, producing stylistically coherent compositions. This study establishes HNote as an effective framework for integrating LLMs with cultural music modeling.

## 1 Introduction

Music generation has long been an important research topic in computational creativity, combining music theory, artificial intelligence, and digital signal processing. With the rapid development of large language models (LLMs), symbolic music modeling has received renewed attention, as LLMs demonstrate strong capabilities in capturing sequential dependencies and stylistic patterns. However, the effectiveness of such models largely depends on the symbolic representation adopted for training.

Traditional music formats such as MIDI [1], MusicXML [2], and ABC Notation [3] provide rich musical semantics but face significant limitations when applied to LLM-based modeling. MIDI focuses on performance instructions, resulting in lengthy and noisy sequences; MusicXML, while comprehensive, is overly verbose and inconsistent across sources; and ABC Notation, though simple and human-readable, lacks strict formatting standards and offers limited expressive power. These factors make traditional formats less ideal for LLM training inputs.

To overcome these challenges, researchers have proposed simplified representations specifically designed for LLMs. For example, YNote employs a fixed-length encoding scheme that balances readability with structural consistency and has shown promising results in generating Jiangnan-style music. However, YNote lacks a precise measure-level alignment mechanism, making it difficult for models to maintain stable rhythmic structures during training.

In this paper, we propose HNote, a novel hexadecimal-based notation system. HNote extends YNote by introducing a fixed 32-unit measure structure and encoding both pitch and duration using a unified hexadecimal vocabulary. This design ensures strict alignment and structural regularity, creating a consistent symbolic space for LLMs. We convert a Jiangnan folk music dataset into HNote and fine-tune LLaMA-3.1, evaluating the results with BLEU and ROUGE metrics. Experimental results show that HNote not only improves the syntactic correctness of generated compositions but also preserves stylistic and structural fidelity, demonstrating its potential as a robust framework for symbolic music generation.

## 2 Music Notation Systems and LLM Fine-Tuning

The effectiveness of music generation tasks is largely determined by the choice of symbolic representation. Each notation system involves trade-offs between informational completeness, structural simplicity, and computational tractability, which in turn shape the efficiency and quality of fine-tuning large language models. In this section, we use the Jiangnan folk tune *"Purple Bamboo Melody" (Zizhudiao)* as an illustrative example and introduce five notation systems: MIDI, MusicXML, ABC Notation, YNote, and HNote. The first three are internationally recognized standards in digital music processing, while the latter two are simplified or extended formats designed specifically for LLM-based modeling. For each representation, we outline its encoding characteristics and review prior studies that employed it for LLM fine-tuning, highlighting both the achievements and challenges. This provides a comparative perspective on the theoretical expressiveness and practical applicability of different notations in symbolic music generation.

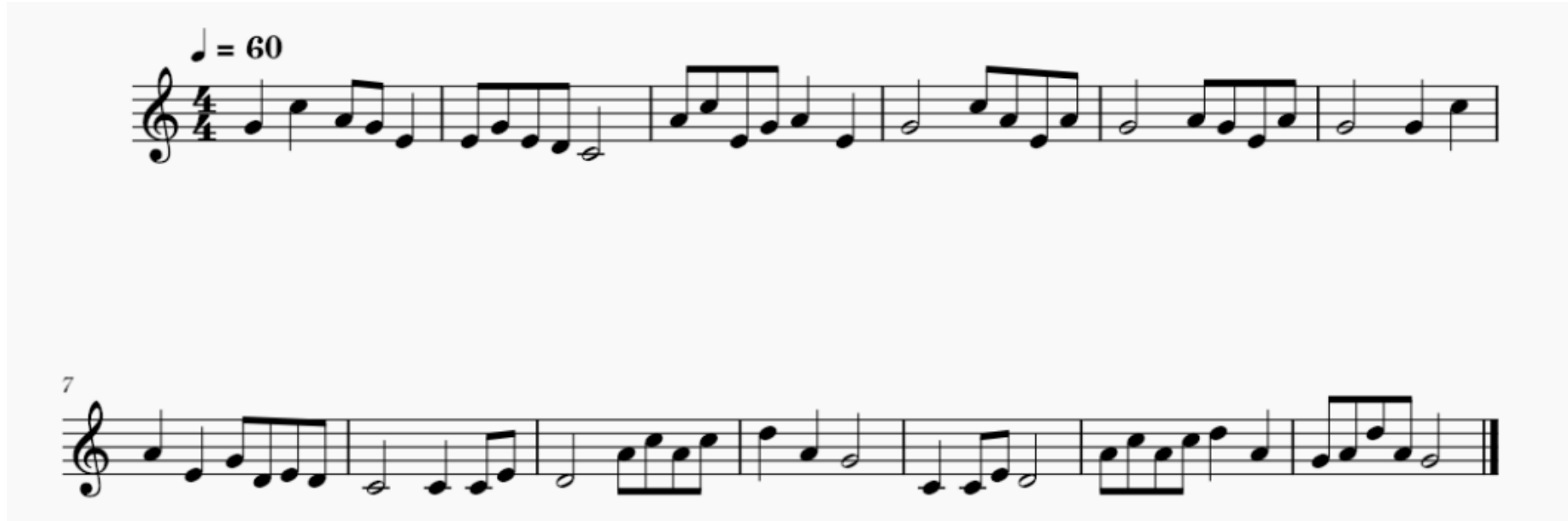

(a) Sheet Music

```
MFile 1 2 120
MTrk
0 Tempo 1000000
0 KeySig 0 major
0 TimeSig 4/4 24 8
0 Meta TrkName "Zizhudiao"

0 On ch=8 n=67 v=90
96 Off ch=8 n=67 v=0
120 On ch=8 n=72 v=90
216 Off ch=8 n=72 v=0
240 On ch=8 n=69 v=90
288 Off ch=8 n=69 v=0
300 On ch=8 n=67 v=90
...
5988 Off ch=8 n=69 v=0
6000 On ch=8 n=67 v=90
6240 Off ch=8 n=67 v=0
6240 Meta TrkEnd
```

(b) MIDI (ASCII Code Representation)

```
<?xml version="1.0" encoding="UTF-8"?>
<score-partwise version="3.1">
  ...
  <measure number="1">
    <attributes>
      <time>
        <beats>4</beats>
        <beat-type>4</beat-type>
      </time>
    </attributes>

    <direction>
      <sound tempo="60"/>
    </direction>

    <note>
      <pitch>
        <step>G</step>
        <octave>4</octave>
      </pitch>
      <duration>2</duration>
      <type>quarter</type>
    </note>
  </measure>
  ...
</score-partwise>
```

(c) MusicXML

```
X:1
T:Zizhudiao
L:1/4
Q:1/4=60
M:4/4
K:C %%stretchlast true
V:1
G c A/G/ E | E/G/E/D/ C2 | A/c/E/G/ A E |
G2 c/A/E/A/ | G2 A/G/E/A/ | G2 G c |
A E G/D/E/D/ | C2 C C/E/ | D2 A/c/A/c/ |
d A G2 | C C/E/ D2 | A/c/A/c/ d A | G/A/d/A/G2|
```

(d) ABC Notation

```
E504G504E504D508E508
G504E508D508E502
E504G504E508D508C504
D504E508G508D54.E508
C504C508A408G44.A408
C504D508E508C502
D54.E508D54.E508
G504A508G508E502
E504G504E508D508C504
D504E508G508D54.E508
C504C508A408G44.A408
C504D508E508C502
```

(e) YNote

```
4FCFCFCFCFCFCFCF54D4D4D4D4D4D4D451D1D1D14FCFCFCF4CCCCCCCCCCCCCCC
4CCCCCCC4FCFCFCF4CCCCCCC4ACACACA48C8C8C8C8C8C8C8C8C8C8C8C8C8C8C8
51D1D1D154D4D4D44CCCCCCC4FCFCFCF51D1D1D1D1D1D1D14CCCCCCCCCCCCCCC
4FCFCFCFCFCFCFCFCFCFCFCFCFCFCFCF54D4D4D451D1D1D14CCCCCCC51D1D1D1
4FCFCFCFCFCFCFCFCFCFCFCFCFCFCFCF51D1D1D14FCFCFCF4CCCCCCC51D1D1D1
4FCFCFCFCFCFCFCFCFCFCFCFCFCFCFCF4FCFCFCFCFCFCFCF54D4D4D4D4D4D4D4
51D1D1D1D1D1D1D14CCCCCCCCCCCCCCC4FCFCFCF4ACACACA4CCCCCCC4ACACACA
48C8C8C8C8C8C8C8C8C8C8C8C8C8C848C8C8C8C8C8C8C848C8C8C84CCCCCCC
4ACACACACACACACACACACACACACACACA51D1D1D154D4D4D451D1D1D154D4D4D4
56D6D6D6D6D6D6D651D1D1D1D1D1D1D14FCFCFCFCFCFCFCFCFCFCFCFCFCFCFCF
48C8C8C8C8C8C848C8C8C84CCCCCCC4ACACACACACACACACACACACACACACACACA
51D1D1D154D4D4D451D1D1D154D4D4D456D6D6D6D6D6D6D651D1D1D1D1D1D1D1
4FCFCFCF51D1D1D156D6D6D651D1D1D14FCFCFCFCFCFCFCFCFCFCFCFCFCFCFCF
```

(f) HNote

Figure 1. ：*Zizhudiao* in Various Music Notations

### 2.1 MIDI

MIDI (Musical Instrument Digital Interface) is a symbolic representation of music that encodes musical information as sequences of discrete events rather than directly storing audio waveforms. MIDI instructions specify pitch, duration, velocity, timing, instrument assignment, and control changes. Due to its structured design and compact file size, MIDI has been widely adopted in digital audio workstations (DAWs), synthesizers, and music education. Compared to waveform data, MIDI is more closely aligned with music theory and compositional structures, making it a common foundation for symbolic music generation and analysis.

In recent years, researchers have attempted to convert MIDI files into textual sequences (e.g., event tokens or simplified symbolic representations) to fine-tune large language models (LLMs) for music generation. However, this approach remains challenging in several aspects. First, the large number of MIDI events produces extremely long token sequences that often exceed the context length of LLMs, making it difficult to capture the global structure of long compositions. Second, inconsistencies and noise in MIDI data—such as incomplete instrument labels or missing control signals—introduce bias into the learning process. Moreover, although generated outputs may be syntactically valid, they often lack harmonic coherence, melodic fluency, or complete formal structure, and their alignment with textual descriptions is frequently imprecise. Finally, handling polyphony and orchestration remains a major difficulty, as current models continue to struggle with multi-instrument arrangements.

For instance, the Text2MIDI model outperformed baseline methods such as MuseCoco in both automatic and human evaluations, demonstrating its ability to better capture the alignment between textual descriptions and symbolic music. Nevertheless, it still encounters difficulties in orchestration control and in maintaining long-term musical structure [4]. Similarly, the MidiCaps dataset provides large-scale aligned MIDI–caption pairs, establishing a valuable resource for fine-tuning LLMs. However, since the captions are typically segment-level descriptions, they fail to represent full-length musical structures, thereby limiting the generative capacity of downstream models [5].

As an illustration, Figure 1_b presents a MIDI file of *Zizhudiao* expressed in ASCII code.

### 2.2 MusicXML

MusicXML is an open standard format designed for the exchange of symbolic music notation. Based on XML, it can encode detailed score information such as pitch, duration, ties, time signatures, tempo markings, dynamics, lyrics, harmony symbols, and ornamentations. Unlike MIDI, which primarily focuses on performance instructions, MusicXML is closer to traditional staff notation and preserves the composer's or arranger's original score intention. Due to its precise alignment with music theory and structured design, MusicXML has been widely adopted in score editing software (e.g., MuseScore, Finale, Sibelius), music analysis, and digital archiving, making it one of the most prevalent formats for music notation exchange.

Although MusicXML provides a comprehensive representation of musical scores and is theoretically well-suited for symbolic input to large language models (LLMs), its application in fine-tuning and generation remains challenging in several aspects. First, MusicXML files are extremely verbose, encoding pitch, rhythm, dynamics, lyrics, ornaments, and layout information. When converted into tokens, this results in very long sequences that often exceed the context length of most LLMs. Second, inconsistencies in encoding practices across different sources lead to heterogeneous and less standardized data, reducing the model's ability to generalize. Moreover, while MusicXML excels at capturing notational accuracy, it lacks the expressive nuances of live performance, such as real-time dynamic shaping or tempo fluctuations. As a result, generated outputs may be structurally valid but musically unnatural. Finally, the scarcity of large-scale MusicXML datasets—particularly those aligned with natural language descriptions—significantly constrains its applicability in text-to-music generation tasks. To illustrate, Figure 1_c presents a MusicXML excerpt of *Zizhudiao*.

### 2.3 ABC Notation

ABC Notation is a text-based music notation system originally proposed by Chris Walshaw in the 1980s, designed to represent melodies using simple ASCII characters to facilitate storage and exchange in plain-text environments. It encodes pitches (letters A–G), octaves (through case shifts or special symbols), note durations (via numeric modifiers), as well as time signatures, tempo markings, and chord indications. Due to its simplicity, ABC Notation has been widely adopted in folk music communities and online platforms, particularly for melody-centric, monophonic music. Compared to formats such as MusicXML or MIDI, ABC Notation is lightweight, human-readable, and easily editable, though it is less suitable for representing complex polyphonic arrangements, detailed dynamics, or nuanced performance instructions.

As a lightweight text-based notation system, ABC Notation has demonstrated unique advantages in fine-tuning large language models for music generation. Breit [6] compared multiple notation formats and found that ABC Notation, owing to its simplicity and readability, enabled more coherent and valid musical outputs than verbose formats such as MusicXML. The study showed that LLMs could generate melodies and chord progressions of “moderate to good” quality, maintaining structural plausibility. However, it also highlighted limitations: ABC Notation primarily supports monophonic melodies, lacking nuanced harmonic representation and expressive dynamics, thereby restricting diversity and richness in the generated music.

Building on this, Yuan et al. [7] introduced ChatMusician, which adopts ABC Notation as its core symbolic representation and leverages continual pre-training and fine-tuning. Their experiments demonstrated significant improvements in format validity, structural repetition control, and human preference ratings compared to GPT-4 and GPT-3.5. Nonetheless, they acknowledged key challenges: the training corpus was heavily biased toward folk music, resulting in stylistic skewness, and the model struggled with deeper harmonic reasoning and polyphonic composition.

Taken together, ABC Notation enables efficient sequence compression and enhances stability in LLM-based music generation, but its limited expressive capacity and dataset bias remain open challenges. Future research should aim to preserve ABC’s simplicity while integrating more diverse corpora and structural enhancements to achieve greater musical complexity and stylistic variety. Figure 1_d provides an example of ABC Notation.

### 2.4 YNote

Lu et al. [8] introduced YNote as a simplified music notation system specifically designed for fine-tuning large language models, with the aim of addressing the complexity and inconsistency of traditional formats such as MIDI, ABC Notation, and MusicXML. YNote employs a fixed four-character encoding scheme to represent both pitch and duration, including dotted notes and ties. Its fixed and uniform structure balances human readability with computational tractability, making it more suitable for tokenization compared to verbose formats like MusicXML and MIDI, which often cause token explosion and structural variability during training.

In their experiments, the authors converted 190 Jiangnan folk music pieces into YNote format and fine-tuned a GPT-2 (124M) model. The results showed that, even with minimal prompt information, the model generated stylistically coherent and structurally reasonable compositions, achieving BLEU scores of up to 0.883 and ROUGE scores of up to 0.766. These outcomes demonstrate the potential of YNote in improving music generation quality and controllability. However, the study also highlighted several limitations. Occasionally, the model produced invalid tokens that required normalization. Additionally, YNote’s four-character constraint cannot fully capture certain complex ornaments and notational details. Finally, the training dataset was limited to Jiangnan-style music, restricting the diversity and generalizability of the generated results. These findings suggest that while YNote provides advantages in simplification and consistency, further research is required to expand its expressive capacity through larger and more diverse datasets as well as enhanced modeling strategies. Figure 1_e provides an example of YNote

# 3 Introduction of HNote

## 3.1 Background

Several symbolic music representations have been widely adopted in previous research on applying large language models (LLMs) to music generation, including MIDI, ABC Notation, and MusicXML. While these formats provide useful structures for encoding musical information, they also present notable limitations when directly applied to LLM training. MIDI primarily focuses on performance control messages and lacks human readability. ABC Notation is compact but suffers from inconsistent formatting due to its flexible syntax and absence of a strict standard. MusicXML, on the other hand, preserves rich musical semantics but is verbose and overly complex for efficient sequence modeling.

The YNote representation [8] was introduced as a simplified and human-readable alternative. YNote improved upon traditional formats by providing a more structured notation system. However, it lacked a fixed alignment mechanism across measures, which made it difficult for LLMs to consistently learn rhythmic structures and maintain temporal regularity during training.

Recent works have nevertheless employed these traditional formats in LLM-based music generation. For example, ChatMusician leverages ABC Notation to treat music as a second language, training LLaMA2 through continual pre-training and fine-tuning on symbolic data [7]. While effective in enabling LLMs to intrinsically understand and generate music, the approach still suffers from the absence of a fixed alignment mechanism for rhythmic structures. Similarly, MuPT introduces SMT-ABC Notation to explicitly address synchronization and measure alignment across tracks, further highlighting that consistent alignment across measures and tracks is essential for preserving rhythmic integrity and structural coherence [9]. In addition, Text2midi-InferAlign explores inference-time alignment to improve the structural consistency of symbolic music generation, demonstrating that alignment constraints can directly enhance output quality, even without modifying the underlying representation [4]. Taken together, these studies underscore that alignment is not a peripheral detail but a central requirement for enabling LLMs to robustly learn and generate structured music.

Building on this foundation, we propose HNote, a novel hexadecimal-based data structure designed to overcome the limitations of YNote. HNote introduces a fixed 32-unit measure structure, ensuring precise alignment of note values and measure boundaries. This design retains the simplicity of YNote while significantly enhancing its suitability for LLM training and music generation, thereby bridging the gap between human readability and LLM learnability.

## 3.2 HNote Design

HNote is a novel data structure designed for representing musical scores. We use two-digit hexadecimal codes as the basic encoding units, which are composed of two key components: Pitch Onset and Note Duration. The former determines the pitch, while the latter specifies the length of the note. In this study, all encodings are expressed in hexadecimal format and enclosed in quotation marks, such as "00", "24", and "7F". In addition, HNote adopts a fixed-length measure structure, where each measure consists of 32 units. This design ensures that different rhythmic values (such as quarter notes, eighth notes, and sixteenth notes) can be precisely aligned within the same framework, thereby maintaining consistency and regularity in the training data, which facilitates large language models in learning musical structures.

#### 3.2.1 Pitch Onset

In HNote, each pitch is represented by a two-digit hexadecimal value ranging from "00" to "7F", covering 128 possible values. Without redefining the rest, the original mapping is "00" = C–1, "01" = C♯–1, with each increment of 1 corresponding to a semitone increase. According to this rule, the lowest note on a standard piano, A0, corresponds to "15", while middle C (C4) corresponds to "3C".

At the same time, pitch in YNote can be directly mapped to HNote Pitch Onset values by their indices. For example, middle C (C4) in YNote corresponds to "3C" in HNote, while the rest is uniformly represented as "00" in both formats.

#### 3.2.2 Note Duration

To distinguish between the onset and continuation of notes, and to avoid ambiguity when representing sustained tones solely with consecutive onsets, HNote defines the range "80" to "FF" as the Note Duration region. This range forms a one-to-one correspondence with "00" to "7F"; for example, the continuation of "00" is "80", the continuation of "01" is "81", and so on.

In HNote, the length of a note is determined by the total number of units consisting of its onset and continuation symbols. For example:

- Whole Note = 32 units (1 onset + 31 duration symbols)
- Dotted Half Note = 24 units
- Half Note = 16 units

Meanwhile, in YNote, note durations were originally represented in beats or fractional units. During conversion, these values are mapped to the corresponding unit counts in HNote and encoded as “onset + continuation.” For instance, a half note represented by index 02 in YNote is converted into 16 units in HNote (including the onset and continuation symbols).

#### 3.2.3 Measure Structure

In this study, HNote adopts a fixed-length measure structure in which each measure contains 32 units. We use 4/4 time as the primary rhythmic basis, where one beat is defined as a quarter note. Since a measure contains four beats, and each quarter note corresponds to 8 units, the total length of a measure is 32 units.

This design offers three major advantages: First, it aligns with fundamental principles of music theory, ensuring rigorous correspondence with traditional time signatures and rhythmic units. Second, all note durations can be precisely represented as integer unit counts (as shown in Figure x). Finally, by fixing each measure to 32 units, all measures in HNote can be exactly aligned at their endpoints. This eliminates alignment issues caused by variable measure lengths and ensures that musical data remain consistent and regular during large language model training, thereby improving the model’s ability to learn rhythmic structures and enhancing the quality of music generation.

As a demonstration, we present *"Purple Bamboo Melody" (Zizhudiao)* as an example of a musical piece converted into the HNote format, as illustrated in Figure 2.

| HNote | YNote | Note | Symbol |
|---|---|---|---|
| 43c3c3c3c3c3c3c3c3c3c3c3c3c3c3c3<br>c3c3c3c3c3c3c3c3c3c3c3c3c3c3c3c3 | G401 | Whole Note | 𝅝 |
| 43c3c3c3c3c3c3c3c3c3c3c3c3c3c3c3 | G402 | Half Note | 𝅗𝅥 |
| 43c3c3c3c3c3c3c3 | G404 | Quarter Note | ♩ |
| 43c3c3c3 | G408 | Eighth Note | ♪ |
| 43c3 | G416 | Sixteenth Note | 𝅘𝅥𝅯 |
| 43 | G432 | Thirty-second Note | 𝅘𝅥𝅰 |

| HNote | YNote | Note | Symbol |
|---|---|---|---|
| 43c3c3c3c3c3c3c3c3c3c3c3c3c3c3c3<br>c3c3c3c3c3c3c3c3 | G42. | Dotted Half Note | 𝅗𝅥. |
| 43c3c3c3c3c3c3c3c3c3c3c3 | G44. | Dotted Quarter Note | ♩. |
| 43c3c3c3c3c3 | G48. | Dotted Eighth Note | ♪. |
| 43c3c3 | G4S. | Dotted Sixteenth Note | 𝅘𝅥𝅯. |

Figure 2. ：Musical Note Symbol Reference Table

# 4 Methodology

## 4.1 Data Preparation

This study employs a dataset of 12,300 Jiangnan-style songs, generated from 123 original Jiangnan folk music pieces through algorithmic composition to capture the stylistic characteristics of traditional Jiangnan melodies. Both the generation algorithm and the original compositions were provided by the authors of YNote to ensure consistency with prior research. All songs were initially encoded in the YNote representation and subsequently converted into the HNote format following the encoding rules described in Section 3. This conversion guarantees consistency and integrity across the dataset, enabling subsequent model training and automatic evaluation to be conducted under a unified and structurally aligned representation.

### 4.2 HNote Conversion

To construct the HNote dataset, we designed an automated conversion pipeline that maps each YNote symbol to its equivalent HNote representation. Specifically:

- Each YNote pitch is mapped to a two-digit hexadecimal code ranging from “00” to “7F,” representing the pitch onset.
- Each YNote note duration is converted into beat units and encoded as a hexadecimal symbol between “80” and “FF,” representing different note lengths.

This design achieves seamless and lossless YNote-to-HNote mapping, avoiding structural ambiguities. Through this process, every piece in the dataset is consistently expressed in HNote, preserving the original melodic and rhythmic information while ensuring that the model learns within a strictly aligned symbolic space.

### 4.3 Model Training

We fine-tuned the LLaMA 3.1 (8B) [10] model as the base large language model, chosen for its balance between performance and hardware requirements, making it suitable for deployment on a single NVIDIA V100 GPU (32GB). Its open-source weights and mature ecosystem also facilitate seamless integration with existing toolchains.

For fine-tuning, we adopted the Unsloth [11] framework, which provides optimized implementations of LoRA (Low-Rank Adaptation). This configuration enables parameter-efficient training while significantly reducing memory usage and training time. On a single V100 GPU, the complete fine-tuning process takes approximately three hours.

During inference, we incorporated soft constraints by specifying the first and last notes of each line as guiding conditions. This design allows the model to better preserve stylistic consistency and structural coherence in the generated compositions, while still maintaining flexibility in the intermediate melodic development.

The overall workflow of the training and inference procedures is illustrated in Fig. 3, which summarizes the end-to-end pipeline from dataset preparation to model evaluation.

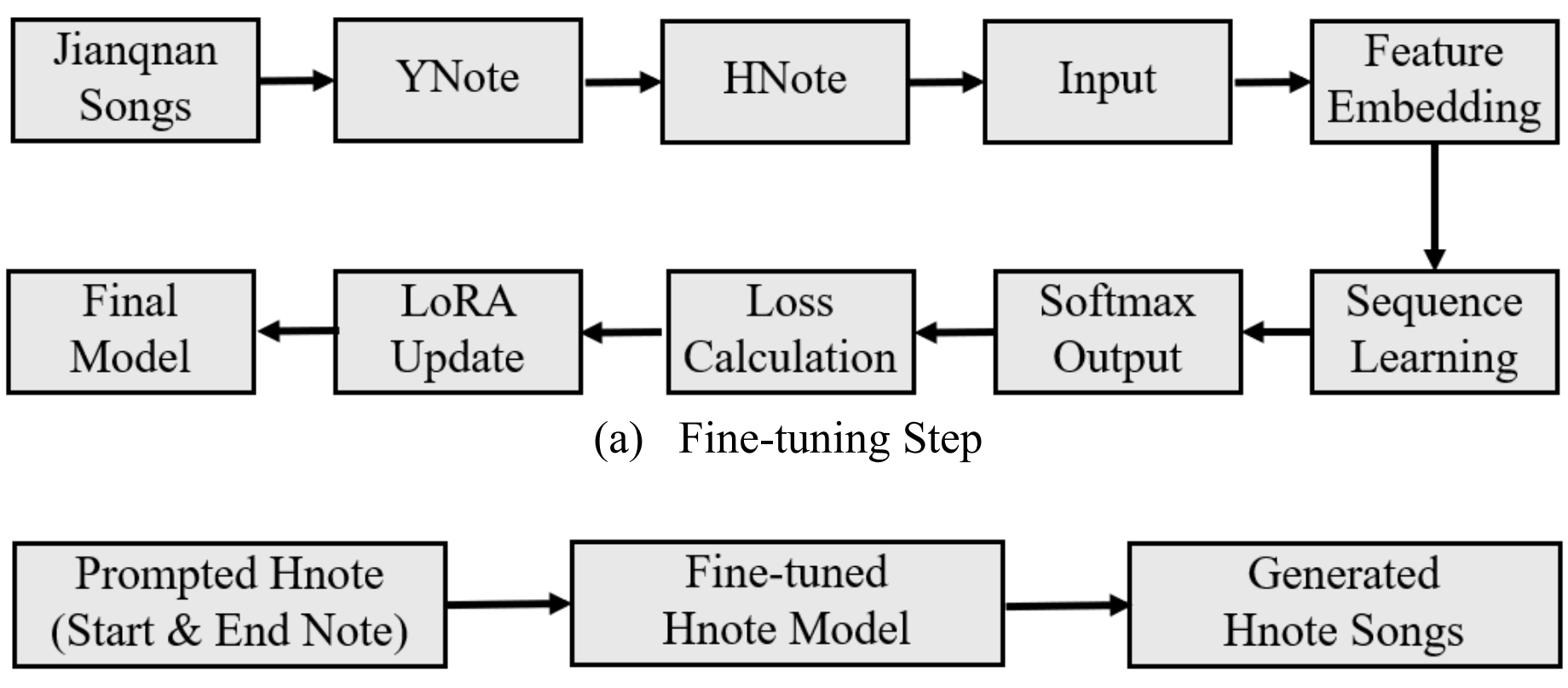

(a) Fine-tuning Step

(b) Inference Step

Figure 3. ：HNote format music generation flow chart

### 4.4 Evaluation

To assess the similarity between generated compositions and reference pieces, we employed three types of automatic evaluation metrics:

- BLEU (Bilingual Evaluation Understudy): Precision-oriented metric that computes the proportion of candidate n-grams appearing in the reference, combined with a brevity penalty to discourage overly short outputs.
- ROUGE-n (Recall-Oriented Understudy for Gisting Evaluation): Recall-oriented metric that measures the proportion of reference n-grams covered by the candidate. In this study, we report ROUGE-1 and ROUGE-2.
- ROUGE-L: Based on the Longest Common Subsequence (LCS), this metric captures global sequence-level similarity.

Together, these metrics reflect different aspects of generation quality: BLEU captures token-level precision, ROUGE-n measures reference coverage, and ROUGE-L evaluates structural fidelity. By incorporating multiple perspectives, our evaluation provides a comprehensive assessment of symbolic and structural consistency in the generated music.

## 5 Experimental Results and Analysis

### 5.1 Correctness of Generation

We first evaluated the syntactic correctness of the generated pieces, defined as whether the outputs strictly adhered to the structural rules of HNote. Out of 1,100 generated pieces, 908 were valid, corresponding to a correctness rate of approximately 82.5%. The remaining errors mainly resulted from incomplete sequences produced during decoding and, in a few cases, missing pitch onset symbols in the HNote encoding. These results demonstrate that HNote provides a highly reliable symbolic framework, with a strong likelihood of generating structurally valid compositions, thereby offering a stable foundation for subsequent automatic evaluation.

### 5.2 Similarity Evaluation (BLEU and ROUGE)

To measure the similarity between generated compositions and reference pieces, we designed two groups of experiments. In the first group, five Jiangnan-style songs were randomly selected from the dataset. For each song, the first and last notes of every line were extracted as prompts to guide generation. The generated outputs were then compared with their corresponding reference pieces, and BLEU-1 to BLEU-4 as well as ROUGE-1, ROUGE-2, and ROUGE-L scores were computed. In the second group, five original Jiangnan songs that were not included among the 123 training songs were selected. The same prompt-and-generation process was applied, and the same evaluation metrics were computed.

As shown in Tables 1 and 2, BLEU and ROUGE scores from both groups were highly consistent across different settings, indicating strong similarity between generated pieces and reference compositions at both the local symbolic level and the global structural level. We attribute this outcome to the structural characteristics of HNote: its regularized encoding ensures that generated and reference sequences remain closely aligned in both length and structure. As a result, precision-oriented BLEU and recall-oriented ROUGE converge to similar values. Furthermore, the high ROUGE-L scores confirm that the generated pieces preserved not only local symbolic overlaps but also global sequential continuity, demonstrating strong fidelity to the style of Jiangnan music.

Table 1. BLEU and ROUGE scores for five songs selected from the dataset

(a) BLEU scores for five songs

| | 1-gram | 2-gram | 3-gram | 4-gram |
|---|---|---|---|---|
| Sample 1 | 0.727 | 0.608 | 0.516 | 0.537 |
| Sample 2 | 0.851 | 0.737 | 0.645 | 0.545 |
| Sample 3 | 0.858 | 0.763 | 0.685 | 0.586 |
| Sample 4 | 0.794 | 0.768 | 0.699 | 0.619 |
| Sample 5 | 0.836 | 0.733 | 0.634 | 0.534 |

(b) ROUGE scores for five songs

| | 1-gram | 2-gram | ROUGE-L |
|---|---|---|---|
| Sample 1 | 0.727 | 0.608 | 0.504 |
| Sample 2 | 0.851 | 0.737 | 0.587 |
| Sample 3 | 0.858 | 0.763 | 0.625 |
| Sample 4 | 0.794 | 0.768 | 0.615 |
| Sample 5 | 0.836 | 0.733 | 0.589 |

Table 2. BLEU and ROUGE scores for five songs selected from the dataset

(a) BLEU scores for five songs

| | 1-gram | 2-gram | 3-gram | 4-gram |
|---|---|---|---|---|
| Sample 1 | 0.746 | 0.624 | 0.519 | 0.387 |
| Sample 2 | 0.837 | 0.735 | 0.668 | 0.604 |
| Sample 3 | 0.843 | 0.753 | 0.645 | 0.474 |
| Sample 4 | 0.691 | 0.589 | 0.503 | 0.426 |
| Sample 5 | 0.758 | 0.561 | 0.449 | 0.368 |

(b) ROUGE scores for five songs

| | 1-gram | 2-gram | ROUGE-L |
|---|---|---|---|
| Sample 1 | 0.746 | 0.624 | 0.535 |
| Sample 2 | 0.837 | 0.735 | 0.628 |
| Sample 3 | 0.843 | 0.753 | 0.613 |
| Sample 4 | 0.691 | 0.589 | 0.475 |
| Sample 5 | 0.758 | 0.561 | 0.547 |

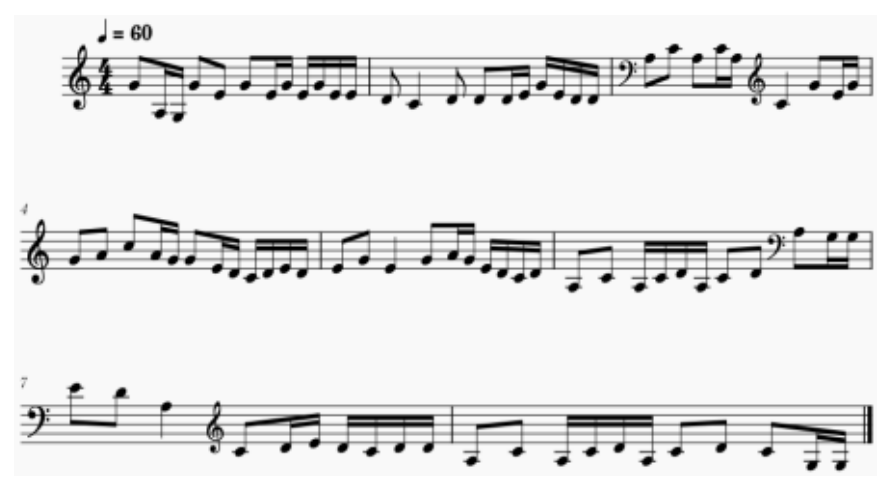

```
Prompt:
starts 4F,4A,45,4F,4C,45,4C,45
ends   4C,4A,4F,4A,4A,43,4A,43
HNote:
4FCFCFCF45C543C34FCFCFCF4CCCCCCC4FCFCFCF4CCC4FCF4CCC4FCF4CCC4CCC
4ACACACA48C8C8C8C8C8C8C84ACACACA4ACACACA4ACA4CCC4FCF4CCC4ACA4ACA
45C5C5C548C8C8C845C5C5C548C845C548C8C8C8C8C8C8C84FCFCFCF4CCC4FCF
4FCFCFCF51D1D1D154D4D4D451D14FCF4FCFCFCF4CCC4ACA48C84ACA4CCC4ACA
4CCCCCCC4FCFCFCF4CCCCCCCCCCCCCCC4FCFCFCF51D14FCF4CCC4ACA48C84ACA
45C5C5C548C8C8C845C548C84ACA45C548C8C8C84ACACACA45C5C5C543C343C3
4CCCCCCC4ACACACA45C5C5C5C5C5C5C548C8C8C84ACA4CCC4ACA48C84ACA4ACA
45C5C5C548C8C8C845C548C84ACA45C548C8C8C84ACACACA48C8C8C843C343C3
```

(a) Sample 1

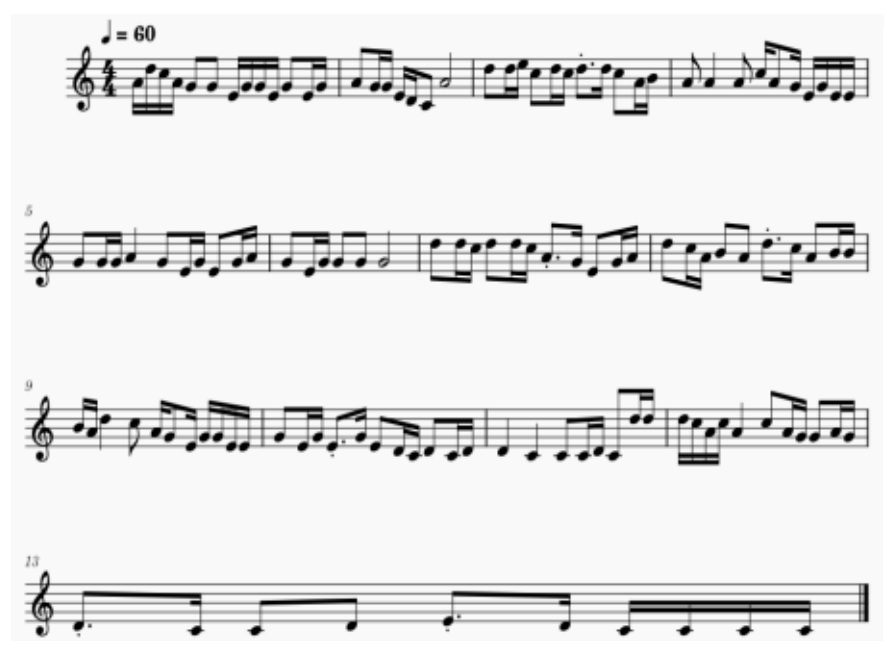

```
Prompt:
starts 51,51,56,51,4F,4F,56,56,53,4F,4A,56,4A
ends   4F,51,53,4C,51,4F,51,53,4C,4A,56,4C,48
HNote:
51D156D654D451D14FCFCFCF4FCFCFCF4CCC4FCF4FCF4CCC4FCFCFCF4CCC4FCF
51D1D1D14FCF4FCF4CCC4ACA48C8C8C851D1D1D1D1D1D1D1D1D1D1D1D1D1D1D1
56D6D6D656D658D854D4D4D456D654D456D6D6D6D6D656D654D4D4D451D153D3
51D1D1D151D1D1D1D1D1D1D151D1D1D154D451D1D1D14FCF4CCC4FCF4CCC4CCC
4FCFCFCF4FCF4FCF51D1D1D1D1D1D1D14FCFCFCF4CCC4FCF4CCCCCCC4FCF51D1
4FCFCFCF4CCC4FCF4FCFCFCF4FCFCFCF4FCFCFCFCFCFCFCFCFCFCFCFCFCFCFCF
56D6D6D656D654D456D6D6D656D654D451D1D1D1D1D14FCF4CCCCCCC4FCF51D1
56D6D6D654D451D153D3D3D351D1D1D156D6D6D6D6D654D451D1D1D153D353D3
53D351D156D6D6D6D6D6D6D654D4D4D451D14FCFCFCF4CCC4FCF4FCF4CCC4CCC
4FCFCFCF4CCC4FCF4CCCCCCCCCCC4FCF4CCCCCCC4ACA48C84ACACACA48C84ACA
4ACACACACACACACA48C8C8C8C8C8C848C8C8C848C84ACA48C8C8C856D656D6
56D654D451D154D451D1D1D1D1D1D154D4D4D451D14FCF4FCFCFCF51D14FCF
4ACACACACACA48C848C8C8C84ACACACA4CCCCCCCCCCC4ACA48C848C848C848C8
```

(b) Sample 2

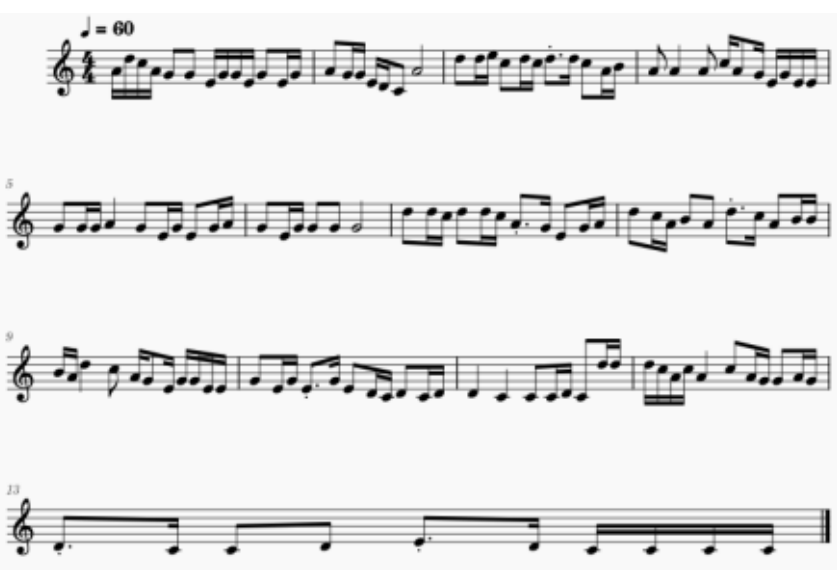

```
Prompt:
starts 54,51,4A,4F,4F,4F,4F,45,48
ends   54,4F,51,4F,4C,51,4F,51,48
HNote:
54D4D4D454D456D654D4D4D454D4D4D451D14FCF4CCC4FCF4FCFCFCFCFCF4CCC
51D1D1D154D451D154D4D4D454D451D14FCFCFCFCFCFCFCFCFCFCFCFCFCFCFCF
4ACACACA4CCC4FCF51D1D1D154D454D454D4D4D451D154D454D4D4D451D151D1
4FCFCFCF51D14FCF4CCCCCCC4FCF51D14FCFCFCFCFCFCFCFCFCFCFCFCFCFCFCF
4FCF51D14FCF4CCC4ACACACA4CCC4FCF4CCC4ACA48C845C548C8C8C84ACA4CCC
4FCFCFCF4FCF4CCC4ACACACA4CCC4FCF4CCCCCCC4ACACACACACACACA48C851D1
4FCFCFCF4CCC4FCF4CCC4FCF51D14FCF4CCCCCCCCCCCCCCC4FCFCFCFCFCFCFCF
45C5C5C548C84ACA4CCCCCCC4FCF4CCC4FCFCFCF51D1D1D154D4D4D451D151D1
48C8C8C8C8C84ACA4CCC4ACA48C8C8C84ACACACACACA48C848C8C8C8C8C8C8C8
```

(c) Sample 3

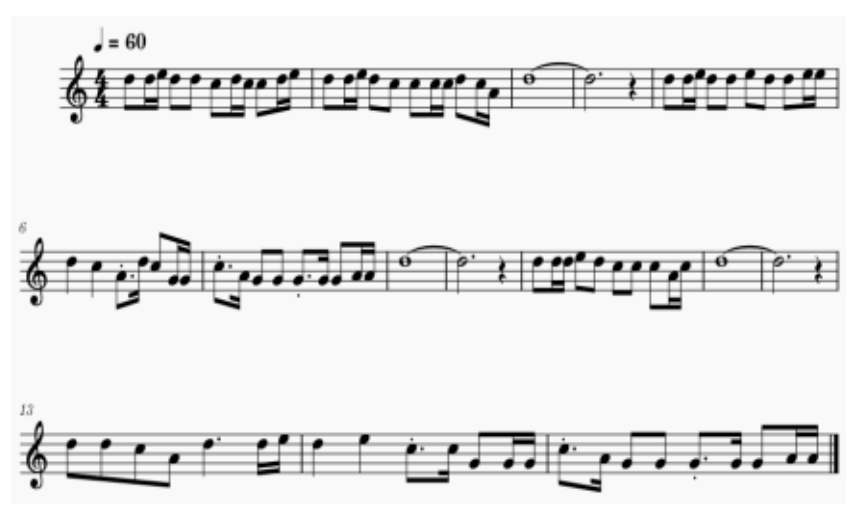

```
Prompt:
starts 56,56,56,56,56,54,56,56,56,56,56,54
ends   58,54,56,58,4F,51,56,54,56,58,4F,51
HNote:
56D6D6D656D658D856D6D6D656D6D6D654D4D4D456D654D454D4D4D456D658D8
56D6D6D656D658D856D6D6D654D4D4D454D4D4D454D454D456D6D6D654D451D1
56D6D6D6D6D6D6D6D6D6D6D6D6D6D6D6D6D6D6D6D6D6D6D6D6D6D6D6D6D6D6D6
56D6D6D656D658D856D6D6D656D6D6D658D8D8D856D6D6D656D6D6D658D858D8
56D6D6D6D6D6D6D654D4D4D4D4D4D4D451D1D1D1D1D156D654D4D4D44FCF4FCF
54D4D4D4D4D451D14FCFCFCF4FCFCFCF4FCFCFCFCFCF4FCF4FCFCFCF51D151D1
56D6D6D6D6D6D6D6D6D6D6D6D6D6D6D6D6D6D6D6D6D6D6D6D6D6D6D6D6D6D6D6
56D6D6D656D656D658D8D8D856D6D6D654D4D4D454D4D4D454D4D4D451D154D4
56D6D6D6D6D6D6D6D6D6D6D6D6D6D6D6D6D6D6D6D6D6D6D6D6D6D6D6D6D6D6D6
56D6D6D656D6D6D654D4D4D451D1D1D156D6D6D6D6D6D6D6D6D6D6D656D658D8
56D6D6D6D6D6D6D658D8D8D8D8D8D8D854D4D4D4D4D454D44FCFCFCF4FCF4FCF
54D4D4D4D4D451D14FCFCFCF4FCFCFCF4FCFCFCFCFCF4FCF4FCFCFCF51D151D1
```

(d) Sample 4

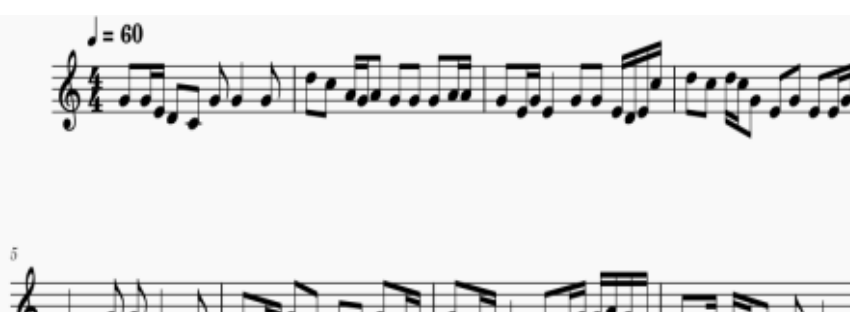

```
Prompt:
starts 4F,56,4F,56,4C,4C,4F,4A
ends   4F,51,54,4F,4C,4A,4C,48
HNote:
4FCFCFCF4FCF4CCC4ACACACA48C8C8C84FCFCFCF4FCFCFCFCFCFCFCF4FCFCFCF
56D6D6D654D4D4D451D14FCF51D1D1D14FCFCFCF4FCFCFCF4FCFCFCF51D151D1
4FCFCFCF4CCC4FCF4CCCCCCCCCCCCCCC4FCFCFCF4FCFCFCF4CCC4ACA4CCC54D4
56D6D6D654D4D4D456D654D44FCFCFCF4CCCCCCC4FCFCFCF4CCCCCCC4CCC4FCF
4CCCCCCCCCCCCCCCCCCCCCCC4FCFCFCF4FCFCFCF4CCCCCCCCCCCCCCC4CCCCCCC
4CCCCCCC4ACA48C84FCFCFCF4CCCCCCC4ACACACA48C8C8C84FCFCFCF4CCC4ACA
4FCFCFCF4CCC4ACA48C8C8C8C8C8C8C84ACACACA4CCC4FCF4FCF51D14FCF4CCC
4ACACACACACA4ACA4CCC4ACA48C8C8C84ACACACA48C8C8C8C8C8C8C8C8C8C8C8
```

(e) Sample 5

Figure 4. ：Music Generation with Fine-Tuned LLaMA-3.1 Using Segment-Initial and Segment-Final Notes as Prompts

## 6 Conclusion

In this study, we proposed HNote as a structured symbolic representation for music generation and evaluated its effectiveness through large language model fine-tuning. By constructing a dataset of 12,300 Jiangnan-style songs, converted from the YNote representation into HNote, we ensured a consistent and unambiguous symbolic space suitable for both training and evaluation. Leveraging LLaMA 3.1 (8B) with parameter-efficient fine-tuning via LoRA, we successfully generated 904 valid HNote sequences.

Our experiments first demonstrated that HNote maintains a high syntactic correctness rate of 82.5%, confirming its reliability as a symbolic framework. To further assess the fidelity of generated music, we conducted similarity evaluations using BLEU and ROUGE metrics on two experimental groups: (1) five songs sampled from the dataset, and (2) five external original Jiangnan songs excluded from the 123 training pieces. Across both groups, BLEU and ROUGE scores were highly consistent, indicating that the generated pieces not only preserved local token-level precision but also maintained global sequential structures. The convergence of precision-oriented BLEU and recall-oriented ROUGE, alongside the high ROUGE-L scores, highlights the structural alignment benefits of HNote and its capacity to capture stylistic and structural fidelity to Jiangnan music.

Overall, this work demonstrates that HNote provides a robust symbolic foundation for music generation with LLMs. It enables consistent evaluation, supports structurally reliable outputs, and facilitates stylistic preservation in generated compositions. For future work, we plan to extend HNote with richer musical annotations, such as chords, dynamics, and tempo variations, in order to capture more expressive dimensions of music. Additionally, incorporating larger training corpora and exploring hybrid evaluation strategies that combine automatic metrics with human listening studies will further strengthen the assessment of generative music systems.

## 7 Future Work

Future work will focus on enriching the expressive capacity of HNote. In particular, we aim to extend the representation to encode chords, accentuation (strong/weak beats), and tempo variations at the measure level, thereby capturing not only pitch and duration but also higher-level musical semantics. Such extensions would enable LLMs to generate music with greater structural richness and stylistic nuance, moving closer to the complexity of human composition.

## Acknowledgement

The authors would like to express their sincere gratitude to National Central University for the support and facilitation that made this research possible. We are also deeply grateful to the National Science and Technology Council (NSTC), Taiwan, for providing financial support under Grant Numbers 113-2622-E-008-020, 113-2221-E-008-095-MY2, and 113-2221-E-008-094-MY2. The generous support from these institutions has been indispensable to the progress and successful completion of this study, for which the authors are truly appreciative.